\documentstyle[12pt, titlepage]{article}
\begin{document}
\begin{titlepage}

\vspace*{5 cm}
\begin{center}
\Large \bf A New Class of Inhomogeneous\\ Cosmological Solutions 
\end{center}

\vspace{1 cm}

\begin{center} 
\normalsize
\bf D. Kamani \dag \mbox{,}  R. Mansouri \S \ddag  \\
\it Preprint 95/7 \\
\normalsize  November 1995
\end{center}

\vspace{1 cm}

\dag Deparment of Physics, Sharif University of Technology, POB. 11365-9161,
 Teheran, Iran. \\
\S On sabbatical leave from Department of Physics, Sharif University of
Technology, Tehran, Iran.\\ 
\ddag Universit\"at Potsdam, Institut f\"ur Mathematik, 
Kosmologie-Gruppe, D-14415 Potsdam, PF 601553, Am Neuen Palais 10, Germany\\ 

\vspace*{1 cm}

\bf Abstract: \normalsize Beginning with a special form of the 
Einstein-Rosen metric,  we find new cosmological solutions of
the Einstein equations, having two hypersurface-orthogonal Killing vectors
, with ideal fluid. The equation of state is in the most cases of the 
form $p = \gamma \rho$. 

\end{titlepage}
\section { Introduction}

\newcommand{\be}{\begin{equation}}
\newcommand{\ee}{\end{equation}}

The actual universe is neither exactly spatially homogeneous nor
exactly isotropic. However, it is much too complicated to
implement these real features of the universe into our models. 
The cosmological principle is the simplifying assumption which 
leads us to a very simple model having only one degree of
freedom: the cosmological scale parameter of FRW-metrics. Now,
the shortcommings of the standard cosmological model is well
known, and even the old-,new-, choatic-,
and extended- inflationary models are not free from problems. A
better understanding of the standard model, and even the
inflationary ones,towards a more realistic model of the universe
is achieved if one could study solutions
without the implementation of cosmological principle.

Anisotropic spatially homogeneous cosmological models have been
studied extensively[1]. Even questions of inflation and bubble
nucleation in anisotropic models have been considered[2].
However, the inhomogeneity is not as easy as anisotropy to
implement. This is why we do not know much about it yet.
Considering the most general inhomogeneous model is yet out of
question, and one should therefore try to understand the
simplest inhomogeneous models. These are models with
one-dimensional inhomogeneity having two space-like commuting
Killing vectors known as orthogonally transitive $G_{2}$
cosmologies [3-5]. Up to now, there are few solutions of this
type known which shows a variety of features[6-10]. The first
class of solutions was given by Wainwright and Goode[6]. The
class of solutions found by Ruiz and Senovilla[8] have quite
different behaviours as far as the structure of singularities
is concerned. There are solutions with a big-bang-like 
singularity, solutions with time-like singularities only in the
Weyl-tensor, solutions with singularities in both the Ricci-
and the Weyl-tensor, and also singularity-free solutions.
Aguirregabia et.al. [9] are interested in inflationary behaviour of
inhomogeneous models with exponential-potential scalar field
as the source, and they have found solutions with multiple
inflationary behaviour for which the deceleration and the
inflationary phases interchange with each other several times 
during the history of the model. Although these models tend to
homogenize, but they never isotropize. 

\section { Metric and field equations}

The simplest one-dimensional inhomogeneity can be formulated
via the Einstein-Rosen metric[4,5]:
\be ds^{2} = - A dt^{2} + B dx^{2} + C(Ddy^{2} + D^{-1}dz^{2}) \ee
where $A$,$B$,$C$,and $D$ are functions of $t$ and $x$. This
metric represents a space-time having two commuting hypersurface
-orthogonal Killing vectors. To simplify the solution all the
authors assume the unknown functions to be separable in $t$
and $x$. Besides this, authors of [6] and [9] assume $D=D(t)$ 
and $C = C(t)$, respectively. Assuming $A=B$, and that the time
part of $A$ be a power of that of the function $G$ we obtain, in
essence, the class of solution of [8].         

We have been able to find some classes of new solutions of the
metric in the following form
\be ds^{2} = -F^{2}dt^{2} + T^{2m}dx^{2} + T.G(T^{n}Wdy^{2} +
 T^{-n}W^{-1}dz^{2}) \ee
Here is $T$ a function of time and $F$,$G$, and $W$ functions
 of  $x$ only. We assume the universe to be filled with perfect 
fluid, having an equation of state to be determined later on.
The Killing vectors are $\frac{\partial}{\partial y}$ and 
$\frac{\partial}{\partial z}$; the unit velocity vector of the
fluid is \be u=-F^{-1}\frac{\partial}{\partial t} \ee  
The energy momentum tensor of perfect fluid is taken to be

\be T_{\mu\nu}= (\rho+p)u_\mu u_\nu + p\eta_{\mu\nu} \ee
To calculate the field equations we use the orthonormal tetrad
$$ 
\omega^{t}=Fdt \quad \mbox{,}\quad \omega^{x}= T^{m} dx 
$$
\be
\omega^{y} = T^{(n+1)/2} \sqrt{GW} dy\quad \mbox{,}\quad
\omega^{z} = T^{(1-n)/2} \sqrt{G/W} dz
\ee 
Now, for the line element (2) and the matter source (4) we
obtain the following non-trivial and independent Einstein field
equations: 
\be
\dot T \left[ -\frac{F'}{F} + \frac{n - 2m + 1}{2}\frac{(\sqrt{GW})'}
{\sqrt{GW}} +
\frac{1 - n - 2m}{2}\frac{ ( \sqrt{GW})'}{ \sqrt{GW}}\right] = 0
 \ee
\be
\frac{n}{F^2} \left[ \left( \frac{ \dot T}{T}\right)^{.}
 + (m+1)\left(\frac{\dot T}{T}\right)^2 \right] 
- T^{-2m}\left[\frac{F'W'}{FW}+\frac{G'W'}{GW}
+ \left(\frac{W'}{W}\right)'\right] = 0
\ee
$$
 \frac{2m-n-1}{2 F^2} \left[
\left(\frac{ \dot T}{T}\right)^{.}+(m+1)\left(\frac{ \dot T}{T}\right)^{2}
 \right] + T^{-2m}[- \left(\frac{F'}{F}\right)' - \left(\frac{F'}{F}\right)^2
$$
\be 
- \left(\frac{W'}{W}\right)^2
+ \frac{W''}{2W}- \frac{1}{2}\left(\frac{G'}{G}\right)'
+ \frac{1}{2}\left(\frac{F'G'}{FG} +
\frac{F'W'}{FW} + \frac{G'W'}{GW}\right)] = 0 
\ee
\be \frac{4m-n^2+1}{4} \frac{1}{F^{2}}\left(\frac{\dot T}{T}\right)^{2} -
T^{-2m} \left(\frac{G''}{G} + \frac{W'^2 }{4W^{2}} 
- \frac{G'^2}{4G^2} \right) = \kappa \rho \ee
\be -\frac{1}{F^{2}}\left[\frac{ n^2+3}{4}
\left( \frac{ \dot T}{T}\right)^{2}
+ \left( \frac{ \dot T}{T}\right)^{.}\right] + 
T^{-2m}\left[\frac{G'^2}{4 G^{2}}-\frac{W'^2}{4 W^2}+
\frac{F'G'}{FG}\right] =
\kappa p \ee
The equation (6) can be integrated to lead
\be F^{2} = G^{1-2m}W^{n}   \ee
In equation (7) the variables can be separated. We call the
separation parameter $\epsilon b^2$, where $\epsilon$ is
choosen to be 0 or $\pm 1$:
\be
\left[ \left( \frac{\dot T}{T}\right)^{.}
+ (m+1)\left( \frac{ \dot T}{T}\right)^{2}\right] T^{2m} =
\frac{ F^{2}}{n}\left[ \frac{F'W'}{FW} + \frac{G'W'}{GW}
+ \left( \frac{W'}{W}\right)' \right] = \epsilon b^{2}
\ee 
Using  equation (11) and defining new variables
\be \alpha =\frac{W'}{W}\quad\mbox{,}\quad
   \beta=\frac{G'}{G} \ee
the field equations (6-8) can be written in the following form
\be \frac{F'}{F}=\frac{n}{2}\alpha+\frac{1-2m}{2}\beta \ee
$$
F^{2} [(1-n)\alpha'+2(m-1)\beta'+\frac{-n^2+n-2}{2} \alpha^{2}+
m(1-2m)\beta^{2} +   
$$
\be
+ \frac{2m(2n-1)-n+3}{2} \alpha \beta ] = -\epsilon (2m-n-1)b^{2} 
\ee
\be
F^{2}\left[\alpha' + \frac{n}{2}\alpha^{2}
+ \frac{3-2m}{2} \alpha \beta\right]=\epsilon nb^{2}  \ee
\be  T^{2m} \left[\left(\frac{\dot T}{T}\right)^{.}+
(m+1) \left(\frac{\dot T}{T}\right)^{2}\right]  
= \epsilon b^{2}\ee 
The four equations (14-17) can be solved to find the unknown
functions $F$, $T$, $G$, and $W$. The equations (9) and (10)
then give pressure and density as functions of $x$ and $t$. It
remains to be verified if there exists an adequate equation of 
state.

The kinematical quantities can now be calculated easily.  The
rotation comes out to be zero. The other non-zero components of
expansion, acceleration, and shear, in tetrad components, are
\be \theta = (m+1)\frac{ \dot T}{FT} \ee
\be  a_{x} = \frac{F'}{T^{m}F}\ee
\be  \sigma_{xx} = \frac{2m-1}{3}\frac{ \dot T}{FT} \ee
\be  \sigma_{yy} = \frac{3n-2m+1}{6}\frac{ \dot T}{FT} \ee
\be  \sigma_{zz} = \frac{1-3n-2m}{6}\frac{ \dot T}{FT} \ee

\section {Solutions}

We differenciate the cases according to the different values of
$\epsilon$.\\

Case I: $\epsilon =0$. \\
Equation (17) can then be integrated immediately to give $T$:
\be T(t) = C_2 \left[C_1 + (m+1)t\right]^{m+1}  \ee
where $C_1$ and $C_2$ are constants of integration. Now assuming
\be G = W^k  \ee with constant $k$, we
  see that $\beta =k \alpha$, where $\beta$ and $\alpha$ are defined in
(13). With this ansatz the equations (15) and (16) can be solved
to give

\be
 k = \frac{n\pm \sqrt{n^{2}+3-4m}}{4m-3} \quad\mbox{,}\quad
n^2-4m+3 \geq 0 
\ee
\be
 W = C'_{2} \left[ C'_{1} + (n+k(3-2m)) x
\right]^{\frac{2}{n+k(3-2m)}} 
\ee
From (24) and (11) the other metric functions are determined.
The remaining field equations (9) and (10) gives $p$ and $\rho$
and the equation of state. It comes out
\be
 \kappa \rho = \kappa p = \frac{4m-n^{2}+1}{4} W^{-[n+k(1-2m)]}
\left(\frac{1}{C_1+(m+1)t}\right)^{2}
\ee
Hence, the fluid obey a stiff equation of state. For the density
to be positive there must be
\be 4m-n^{2}+1 > 0  \ee
This gives together with (25)
\be  4 > 4m-n^{2}+1 > 0   \ee
The case $m=-1$, which would lead to constant $T$, is now
automatically excluded.
Obviously the zeros of the functions $T(t)$ and $W(x)$ leads to 
singularities. The former is a big-bang-like singularity. The latter
one, which is generic in most of inhomogenous solutions yet found
has not been studied throughly. To study such type of singularities, which
we call them 'wall-like singularity', it is necessary to pick up some
tractable special solutions suitable for physical interpretations.\\   

Case II: $\epsilon = +1$.\\

 The integration of (17) in the following cases is straightforward:
\be m = 0 \quad\mbox{:}\quad
 T(t) = C_1 e^{bt} + C_2 e^{-bt} \ee 
\be
m = +1 \quad\mbox{:}\quad T(t) = \pm \sqrt{ b^{2}t^{2}+C_{1}t+C_{2}}
 \ee
\be
m = -1 \quad\mbox{:}\quad T(t) = \frac{C_1}{b \sinh(C_2 \pm C_1 t)} 
\ee
Now, assuming
\be  T(t) = [U(t)]^{\lambda} \ee
with constant $\lambda$, the equation (17) can be integrated  in general to
give 
\be \dot U^2 = C_{1} + b^2 (m+1)^2  U^{\frac{2}{1+m}}  \ee
For the integration to be possible we have choosen $\lambda$ to be
equal $\frac{1}{m+1}$. $C_{1}$ is again a constant of integration.
For $C_{1}=0$, the equation (34) can be integrated. We obtain
\be T(t) = (C\pm mbt)^{\frac{1}{m}} \quad \mbox{with}
\quad m \not= 0,-1  \ee
Note however, that for $m = +1$ this solution is a special case
of (31). Now we find two classes of solutions for the
$x$-dependent functions.\\

Case IIa: Let
\be W = A e^{qx} \ee
\be G = B e^{sx} \ee
where $A,B,q$ and $s$ are constant. From (14-16) we then find
\be F = \mu e^{[nq+(1-2m)s] x/2} \ee
\be \mu^{2} e^{[nq+(1-2m)s] x} 
\left(\frac{n}{2}q^{2}+\frac{3-2m}{2}qs\right) = nb^{2} \ee
$$
\mu^{2} e^{[nq+(1-2m)s] x}
[\frac{-n^{2}+n-2}{2}q^{2}+m(1-2m)s^{2} + 
$$
\be
+\frac{2m(2n-1)-n+3}{2} qs] = -b^{2}(2m-n-1)
 \ee
From (39) and (40) we must have
\be nq+(1-2m)s=0 \ee
The above equations finally lead to
\be \frac{n}{2}q^{2}+\frac{3-2m}{2}qs = nb^{2}/\mu^{2} \ee
$$
\frac{-n^{2}+n-2}{2}q^{2}+m(1-2m)s^{2}+\frac{2m(2n-1)-n+3}{2} qs =
$$
\be
= -(2m-n-1)b^{2}/\mu^{2} 
\ee
The parameters $q,s$ and $\mu$ are obtained from (41-43). For
the unknown functions we get
\be
F = \mu \quad\mbox{,}\quad \mu = A^{n/2}B^{(1-2m)/2}
 \ee
\be W = A e^{\pm \frac{b}{\mu}\sqrt{2m-1}x} \ee
\be G = B e^{\pm\frac{nb}{\mu}\frac{1}{\sqrt{2m-1}}x} \ee
As we are working with real metric, the above solutions are
restricted to \\
 $m>1/2$. Hence in calculating $\rho $ and $p$ we
have just the choise of (31) and (35) for $T(t)$. Taking the general choice
(35), we obtain from (9,10):
\be
 \kappa \rho = \frac{b^2}{2\mu^2}\frac{(m+1)(2m-n^2-1)}{2m-1}\frac{1}{
(C_2 \pm mbt)^2} \ee
\be \kappa p = \frac{b^2}{2\mu^2}\frac{(m-1)(2m-n^2-1)}{2m-1}\frac{1}
{(C_2 \pm mbt)^2}
 \ee
with $m \geq 1$ and $n^2 \leq 2m-1$. The equation of state comes out to be
\be p = \gamma \rho \quad\mbox{,}\quad \gamma = \frac{m-1}{m+1}\quad
\mbox{,}\quad 0 \leq\gamma<1  \ee
Taking the choice (31) for $m=1$, we obtain
$$
 \kappa \rho = \frac{b^2}{\mu^2}(1-n^2)\times
$$
\be
\times \left[b^2t^2 + C_1 t + 
\frac{(5-n^2)C_1^2 - 4(3n^2+1)C_2 b^2}{16b^2 (1-n^2)} \right]
\left( b^2t^2 + C_1t + C_2 \right)^{-2}
\ee
\be
 \kappa p = \frac{5-n^2}{16\mu^2} \left({C_1}^2 - 4b^2 C_2 \right)
\left( b^2t^2 + C_1 t + C_2 \right)^{-2}
\ee
The proportionality factor between pressure and density is now a function
of time. Looking for a simple equation of state of the form
\be
p = \gamma\rho
\ee
with constant $\gamma$, we are led to the following conditions
$$
C_1 = 2bC \quad\mbox{,}\quad C_2 = C^2
$$
This leads to the following equation of state:
\be
p = 0 \quad\mbox{,}\quad \kappa\rho = \frac{b^2(1-n^2)}{\mu^2}
\frac{1}{(bt + C^2)^2} 
\ee 
This correspond to the solutions (47, 48) for $m = 1$. Hence, we
have a dust solution with a standard big-bang-like singularity.\\

Case IIb: Let
\be \alpha = \frac{D}{F(x)} \quad \mbox{,} \quad \beta = \frac{E}{F(x)} \ee
with constant $D$ and $E$. Substituting in the equations
(14-16) we obtain
\be F(x) = k+ A_{0}x \ee
\be W(x) = k_{1} (k+A_{0} x)^{D/A_{0}} \ee
\be G(x) = k_{2} (k+A_{0} x)^{E/A_{0}} \ee
where
\be A_{0} = \frac{1}{2} nD + \frac{1}{2} (3-2mE) - \frac{nb^{2}}{D}
\ee 
The following consistency relations between  constants must also
hold: 
\be k_{2} = k_{1}^{\frac{n}{2m-1}} \ee
\be ED = nb^{2}  \ee
\be -D^{2}+(3-4m) E^{2}+2nDE+2n(m-1) b^{2} \frac{E}{D}+
(2m-n^{2}-1)b^{2} = 0 \ee
There are two solutions of (60, 61):
\be D^{2} = n^{2}b^{2} \ee
\be D^{2} = (2m-1)b^{2} \ee
Choosing (63) we obtain from (58) that $A_0 = 0$, which is unacceptable
due to the relations (56) and (57). Therefore we try (62). Using (35)
for $T(t)$, it comes out
\be
\kappa \rho = \kappa p = 0
\ee
Now we have still the choices (30-32) at our disposal. The case $m=0$ leads
to the following relations for pressure and density:
\be
\kappa\rho = \kappa p = \frac{(n^2-1)b^2}{(k + A_0 x)^2} C_1 C_2
\left(C_1 e^{bt} + C_2 e^{-bt} \right)^{-2}
\ee
The condition for the positivity of the density is
\be
( n^2-1) C_1 C_2 > 0
\ee
The case $m=-1$ leads to negative density. But for $m=1$ we get
$$
\kappa \rho = \kappa p =
$$
\be
=\frac{(n^2-5)b^2}{4} \left[ 1-\frac{1}{b^2}\frac{(b^2t+C_1/2)^2}{
b^2t^2 + C_1 t + C_2}\right]\frac{1}{(k + A_0 x)^2 (b^2 t^2 + C_1 t +
C_2)}
\ee
The equation of state is again in the stiff form.\\  

Case III: 
 $\epsilon=-1$.\\

Solutions for this case can be find using again
the forms $$W = A e^{qx}$$and$$G = B e^{sx}$$
This choice leads to 
\be  W = A e^{\pm \frac{b}{\mu} \sqrt{1-2m} x}  \ee
\be G = B e^{\mp \frac{nb}{\mu}\frac{1}{\sqrt{1-2m}}x} 
 \ee
\be F = \mu \quad\mbox{,}\quad \mu = A^{n/2} B^{(1-2m)/2} \quad
\mbox{,}\quad  m < 1/2
\ee
The equation (17) leads to 
\be
T(t) = (C \pm imbt)^{\frac{1}{m}} \qquad m \not= 0
\ee
For $m = 0$ one gets a sinosoidal function which leads to violation of energy
condition and is therefore unphysical. We obtain real values of (71) for
\be 
m = \frac{1}{4l} \quad\mbox{,}\quad l\in Z - {0}  \ee
this leads to
\be
T(t) = (\frac{bt}{4l})^{4l}
\ee
Pressure and density are calculated from (9) and (10):
\be
\kappa\rho = \frac{4l^2}{\mu^2}\left[ \frac{3}{2l}-n^2(\frac{8l-1}{2l-1})
\right] \frac{1}{t^2}
\ee
\be
\kappa p= \frac{4l^2}{\mu^2}\left[\frac{3}{2l}-4 + \frac{n^2}{2l-1}\right]
\frac{1}{t^2}
\ee
with the equation of state in the form $p = \gamma\rho$. The positivity
of energy leads to $l > 0$; but the energy condition $0\leq p \leq \rho$
leaves no choise for $n$ and $l$.
Seeking another solution along the lines of the case IIb leads to complex
metric functions.\\

\section{Discussion}

The solutions we have found shows very diverse singularity behaviours.
 There are
some with  big-bang- and wall-like singularities, such as (27) and (67).
Others show just big-bang-like singularities, such as (47,48), (50,51,53),
 and (74,75). The class of solutions leading to (65) is peculiar. It shows
no big-bang singularity. In fact, there are some solutions with
 no singularity in time.
 However, it has a wall-like singularity. The nature of all these
 singularities should be studied very carefully. One has to consider
special solutions and look for any physical interpretations of them.
We are considering some interesting solutions of this class and
those found by other authors, which will be subject of other publications. 

\vspace{2 cm}

\normalsize\bf Acknowledgments \\
\normalsize
One of the authors(R. M.) would like to thank Dr. habil. H.- J. Schmidt 
for hospitality at the Kosmologie-Gruppe, Institut f\"ur Mathematik,
Universit\"at Potsdam, and Alexander v. Humbolt Stiftung for a
research fellowship.  \\

\end{document}